\documentclass[journal,twoside,web]{ieeecolor}
\usepackage{jsen}
\usepackage{cite}
\usepackage{amsmath,amssymb,amsfonts}
\usepackage{graphicx}
\usepackage{textcomp}
\usepackage{wrapfig}

\usepackage{algorithm}
\usepackage{algorithmic}
\makeatletter
\newcommand{\removelatexerror}{\let\@latex@error\@gobble}
\makeatother

\usepackage{amsmath}
\usepackage{diagbox}
\usepackage{multirow}
\usepackage{booktabs}

\def\BibTeX{{\rm B\kern-.05em{\sc i\kern-.025em b}\kern-.08em
    T\kern-.1667em\lower.7ex\hbox{E}\kern-.125emX}}
\definecolor{abstractbg}{rgb}{0.89804,0.94510,0.83137}
\begin{document}
\title{sEMG-based Fine-grained Gesture Recognition via Improved LightGBM Model}
\author{Xiupeng Qiao, Zekun Chen, Shili Liang
\thanks{This work is supported by the Natural Science Foundation of Jilin Province [grant number YDZJ202201ZYTS506. (Corresponding authors: Shili Liang)}
\thanks{Xiupeng Qiao, Zekun Chen, Shili Liang are with the School of
	Physics, Northeast Normal University, Changchun 130022,
	China (e-mail: 1603621303@qq.com, 229800776@qq.com, lsl@nenu.edu.cn.)}}

\IEEEtitleabstractindextext{%
\fcolorbox{abstractbg}{abstractbg}{%
\begin{minipage}{\textwidth}%
\begin{wrapfigure}[12]{r}{3in}%
\includegraphics[width=3in]{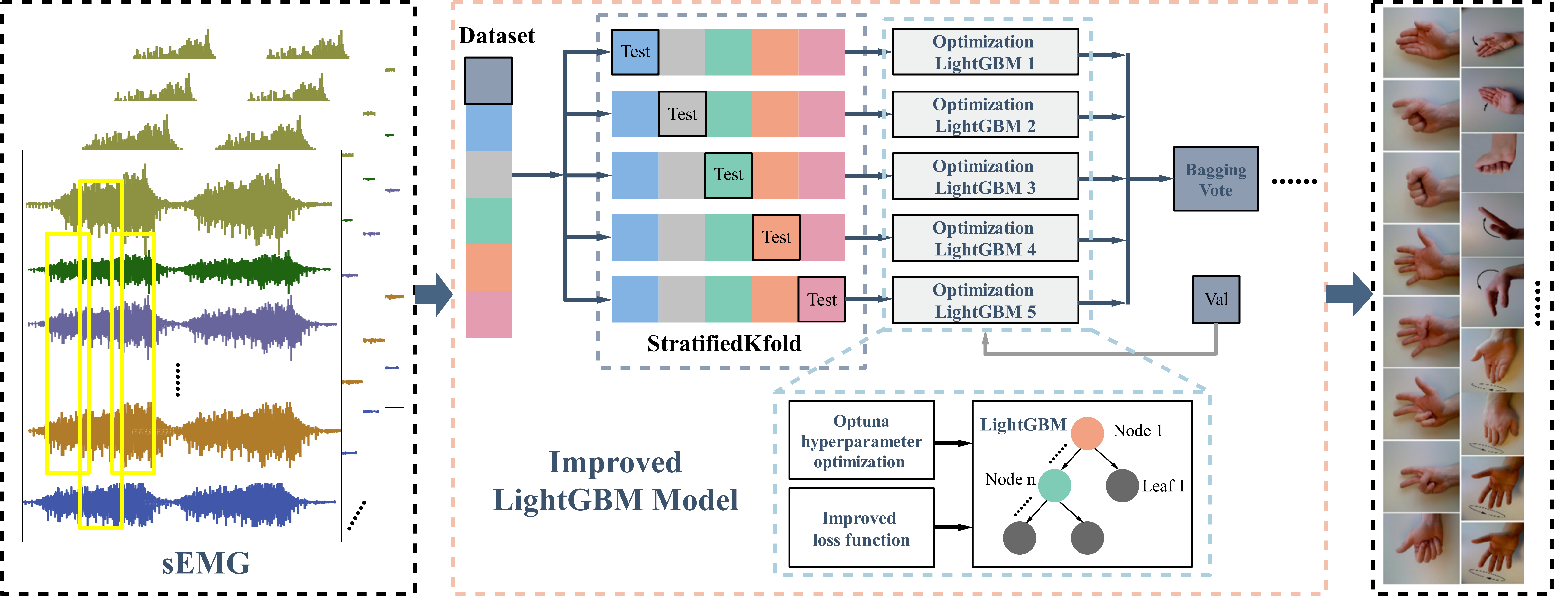}%
\end{wrapfigure}%
\begin{abstract}
Surface electromyogram (sEMG), as a bioelectrical signal reflecting the activity of human muscles, has a wide range of applications in the control of prosthetics, human-computer interaction and so on. However, the existing recognition methods are all discrete actions, that is, every time an action is executed, it is necessary to restore the resting state before the next action, and it is unable to effectively recognize the gestures of continuous actions. To solve this problem, this paper proposes an improved fine gesture recognition model based on LightGBM algorithm. A sliding window sample segmentation scheme is adopted to replace active segment detection, and a series of innovative schemes such as improved loss function, Optuna hyperparameter search and Bagging integration are adopted to optimize LightGBM model and realize gesture recognition of continuous active segment signals. In order to verify the effectiveness of the proposed algorithm, we used the NinaproDB7 dataset to design the normal data recognition experiment and the disabled data transfer experiment. The results showed that the recognition rate of the proposed model was 89.72\% higher than that of the optimal model Bi-ConvGRU for 18 gesture recognition tasks in the open data set, it reached 90.28\%. Compared with the scheme directly trained on small sample data, the recognition rate of transfer learning was significantly improved from 60.35\% to 78.54\%, effectively solving the problem of insufficient data, and proving the applicability and advantages of transfer learning in fine gesture recognition tasks for disabled people.
\end{abstract}

\end{minipage}}}

\maketitle

\section{Introduction}
\label{sec:introduction}
\IEEEPARstart{S}{urface} electromyographic is a weak electrical signal generated by human muscles during contraction. It can not only map human neuromuscular activities, but also be used as an important means to convey human intentions. With the continuous innovation of science and technology, the scope of application of surface EMG has been gradually broadened, and now it has been involved in many fields such as motion analysis, rehabilitation treatment, human-computer interaction and so on.

\begin{figure*}[!t]
	\centering
	\includegraphics[width=0.9 \textwidth]{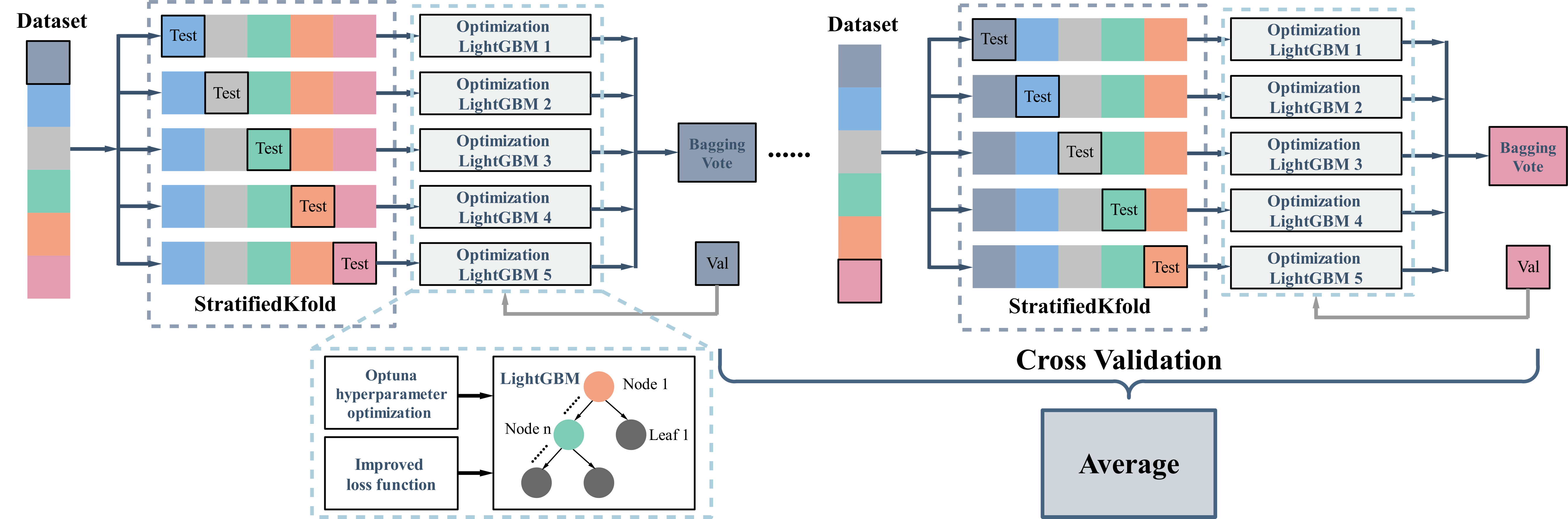}
	\caption{Framework of movements classification based on inproved LightGBM.}
	\label{fig1}       
\end{figure*}

Due to the rapid development of machine learning, various kinds of machine learning algorithms are widely used in the task of gesture recognition of surface EMG. At the very beginning, LDA\cite{ur2017novel}, SVM\cite{al2013classification}, KNN\cite{narayan2021semg} and other methods have achieved initial success in this field. Under the condition of large training set samples, these hand motion recognition models based on machine learning have shown excellent accuracy and robustness of gesture recognition \cite{gozzi2022xai}. (1) Based on the LDA model, Duan et al. \cite{duan2018gesture} extracted the RMS ratio as a time-domain feature and achieved an average recognition rate of 91.7\% in nine gesture classification tasks of three-channel sEMG data. This allowed the machine learning algorithm to achieve initial success in the field of gesture recognition, but the experiment did not fully consider the non-ideal case of sEMG. On the other hand, Naik et al. \cite{naik2014classification} firstly proposed a scheme of using principal component analysis (PCA) and independent component analysis (ICA) to reduce sEMG data and blind source decomposition, and then using LDA algorithm for gesture recognition. This scheme obtains sEMG features with higher signal-to-noise ratio, and the accuracy of gesture recognition reaches 90\%. (2) Based on KNN algorithm, Narayan et al. \cite{narayan2021semg} chose the method of discrete wavelet transform for denoising in the data pre-processing stage, then extracted the time-domain features and time-frequency domain features in the feature engineering stage, and demonstrated the influence of different features on six gesture recognition tasks through experiments, the results show that appropriate features can greatly improve the accuracy of gesture recognition. In order to further solve the problem of poor real-time prediction in gesture recognition tasks, Marco et al. \cite{benalcazar2017real} used the Myo armband as the sEMG data acquisition device, and adopted the sEMG data activity segment detection scheme to ensure the signal instantness. (3) SVM-based gesture recognition studies mostly focus on upper arm movements, while the classification of fine movements of single fingers is relatively rare. Chen et al. \cite{chen2013pattern} extracted time-domain features, spectral power amplitude features and correlation coefficient class features between channel signals as feature sets, and used Multi-core Learning Support Vector Machine (MKL-SVM) for pattern recognition of finger-like actions. The research has achieved remarkable results with recognition accuracy up to 97.93\%, opening up new directions for flexibility and accuracy in hand motion recognition. Due to the extreme flexibility of the fingers, the number of movements will inevitably increase during the acquisition process, resulting in motion artifacts \cite{karacan2023estimating}, which pose a serious challenge for accurate classification.

However, these machine learning methods rely heavily on feature engineering, and the selection and extraction of feature information in sEMG often requires professional knowledge and experience \cite{shao2020single}. In addition, factors such as the number and location of electrodes during sEMG data acquisition \cite{chen2021surface} largely determine the performance of the gesture recognition model. When faced with sEMG data of new subjects or more types of gestures, sEMG gesture recognition based on machine learning algorithm has the problems of limited generalization ability and excessive algorithm complexity, which affect the intelligent development of sEMG gesture recognition to a certain extent. In contrast, neural networks show a broader application prospect in sEMG gesture recognition tasks.

It should be noted that the training of gesture recognition model based on neural network requires a large amount of data \cite{luu2021deep}, and the model's interpretability is poor, which will lead to relatively weak stability and migration ability of the model. In order to address the problem of declining classification accuracy during feature extraction, Ding et al. \cite{ding2018semg} proposed a parallel multi-scale convolutional gesture recognition system. They used a larger kernel filter to conduct classification experiments on the NinaPro database \cite{atzori2015ninapro}, and the results showed that the recognition accuracy of the model was significantly improved. However, this method also leads to excessive complexity of the model, which further leads to overfitting of the training data and poor robustness. Unlike CNN, Recurrent Neural Network (RNN) regards sEMG as time series data and can capture information from adjacent inputs and extract time features of sEMG for gesture recognition \cite{xiong2021deep}. Zhang et al. \cite{zhang2020novel} proposed a novel RNN model, which was able to encode and train the pre-sampling time step of EMG, achieving a classification accuracy of 89.6\%. In spite of this, RNN still has the problem that it cannot effectively remember the early sEMG signal when making prediction, and it is prone to the situation that the gradient disappears during training, so it cannot complete the task of gesture recognition through sEMG signal. In order to improve this problem, various studies introduced the Long Short-Term Memory network (LSTM) \cite{graves2012long} to replace the RNN network and control the long-term memory and short-term memory of information through its internal "gate" mechanism, thus avoiding the occurrence of gradient disappearance to a certain extent. After comparing various gesture recognition models, Samadani et al. \cite{samadani2018gated} found that the Bi-LSTM model was the optimal model on the NinaPro DB2 dataset, with an accuracy rate of 86.7\%.

To sum up, most gesture recognition systems currently classify gestures based on static or quasi-static gestures, while recognition of continuous dynamic gestures is relatively rare. Continuous dynamic gestures are more difficult to identify because they involve more muscle synergies and spatio-temporal variations \cite{yang2017semg}. However, in practical applications, continuous dynamic gestures are often more practical. In addition, how to enable amputees to use sEMG signals of residual arms to achieve fine gesture classification is also an urgent problem to be solved. Since the physiological structure of amputees is significantly different from that of normal people, the muscle distribution and nerve conduction path of the residual arms may be changed \cite{cote2017transfer}. Therefore, the traditional gesture recognition algorithm may not be directly applied to amputees, and it is necessary to optimize and improve the algorithm according to the special physiological conditions of amputees to improve the accuracy and robustness of gesture recognition. At the same time, it is necessary to work closely with medical rehabilitation institutions to carry out a large number of clinical trials and validation work to ensure the safety and effectiveness of the algorithm.

Motivated by this, in this paper, an improved fine gesture recognition model based on LightGBM algorithm is proposed. First, precision monitoring is designed as early-stop condition to prevent overfitting of LightGBM model, and Optuna is used to search for optimal hyperparameters to improve the model's capability. Secondly, StraitiedKold combined with bagging is used to offset the effect of uneven distribution of the training set and test set. Then, the basic cross-entropy loss function is improved to make the model pay more attention to the recognition of difficult gestures. Finally, the accuracy of the improved LightGBM model is compared with that of existing studies, and the effectiveness, robustness and generalization of the proposed algorithm are verified. The framework is shown in Fig. \ref{fig1}.

The remaining structure of this article is as follows. In Section \ref{section2}, the proposed framework is described in detail. Section \ref{section3} gives the experimental results and discussion. Section \ref{section3} gives the conclusion and the direction of future work.

\section{Methods}
\label{section2}
\subsection{sEMG Acquisition}
\begin{figure}[!t]
	\centering
	\includegraphics[width=0.4 \textwidth]{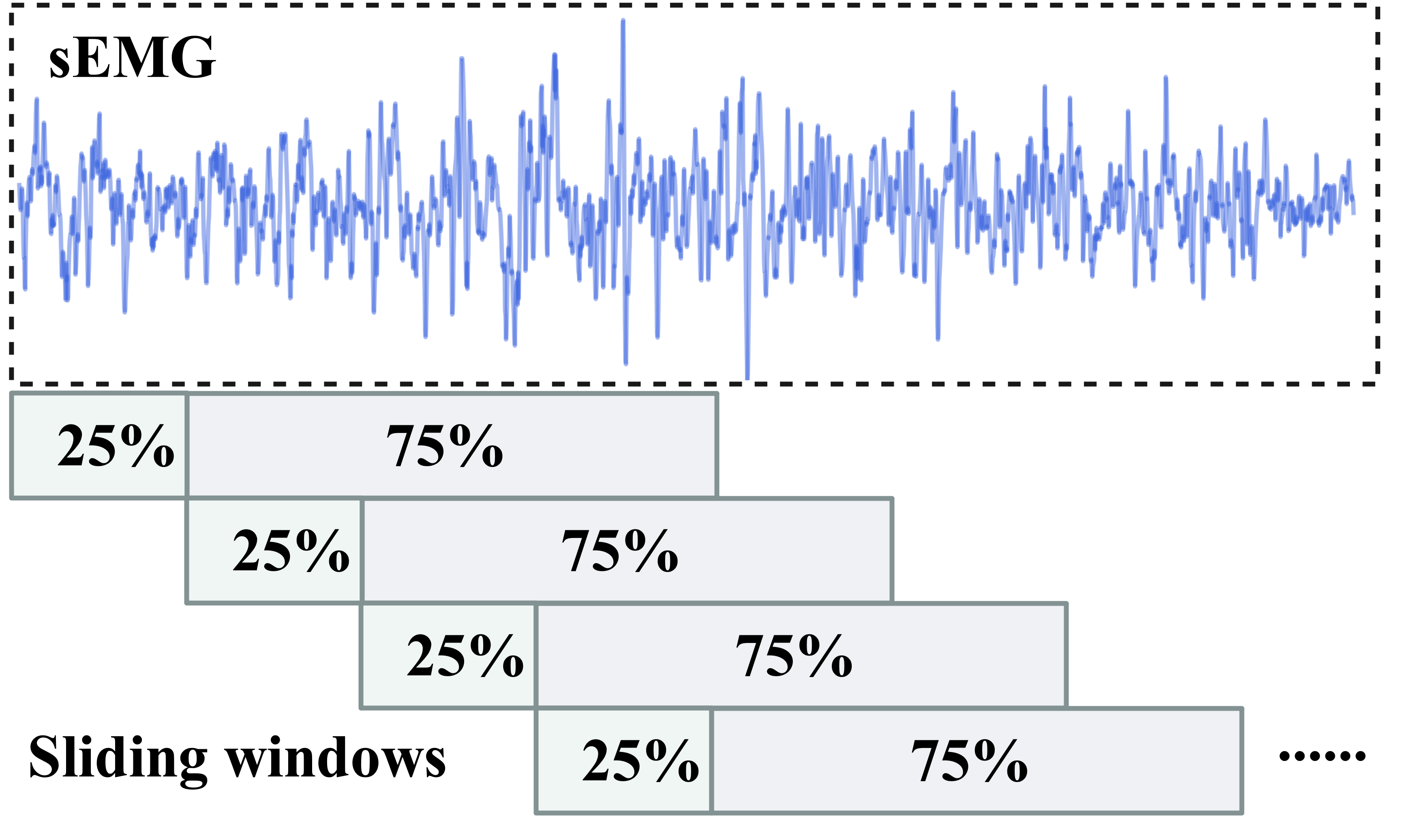}
	\caption{Schematic diagram of the segmentation sample}
	\label{fig2}       
\end{figure}

The data are derived from the NinaproDB7 dataset \cite{atzori2015ninapro}, which was developed and published by the NinaPro project team at the University of Applied Sciences (HES-SO) in Switzerland to study gesture recognition and human-computer interaction. The dataset collected EMG signals from 22 subjects (20 normal subjects and 2 disabled subjects) and included three major groups of gesture movements, of which the second group of 18 finger movements and wrist movements were the most studied. We choose the method of sliding window to sample the original signal. In order to improve the practicability and robustness of the final model, we did not choose to divide the training set and the test set by scrambling all samples after segmentation, but randomly grouped the number of actions to avoid data leakage. The specific segmentation method is shown in Fig. \ref{fig2}. When sliding the window, 75\% time-domain window stack is adopted. If the adjacent time Windows in the time-domain are divided into the training set and the test set respectively, 75\% of the sample points of the test set have been learned by the model in the training set, so the model evaluation index obtained is distorted and cannot be used as the evaluation standard for gesture recognition rate in reality. So the six movements of all subjects were randomly divided into three groups with two movements in each group. Each group was used as a separate test set, and the other two groups were used as a training set for cross-verification experiments. The average value of the experimental results of the three modeling sessions was taken. The grouping of random groups is shown in Table \ref{tab:1}. In the subsequent experiments, this grouping method is carried out three times to improve the credibility of the model through this cross-verification method, so as to ensure that the model does not simply have a high classification performance for a certain action data.

\subsection{Pre-Processing}
\begin{figure}[!t]
	\centering
	\includegraphics[width=0.16 \textwidth]{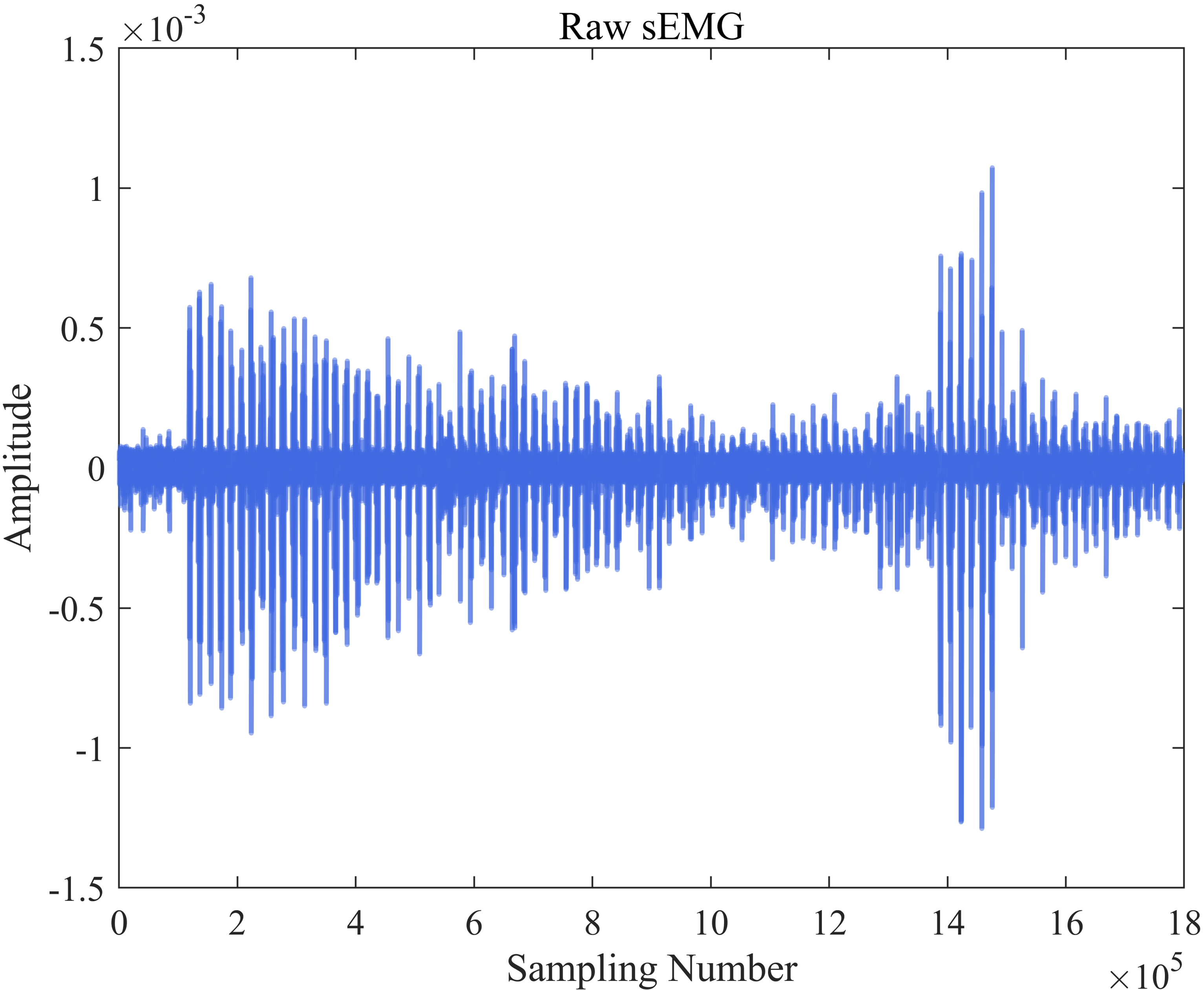}\includegraphics[width=0.16 \textwidth]{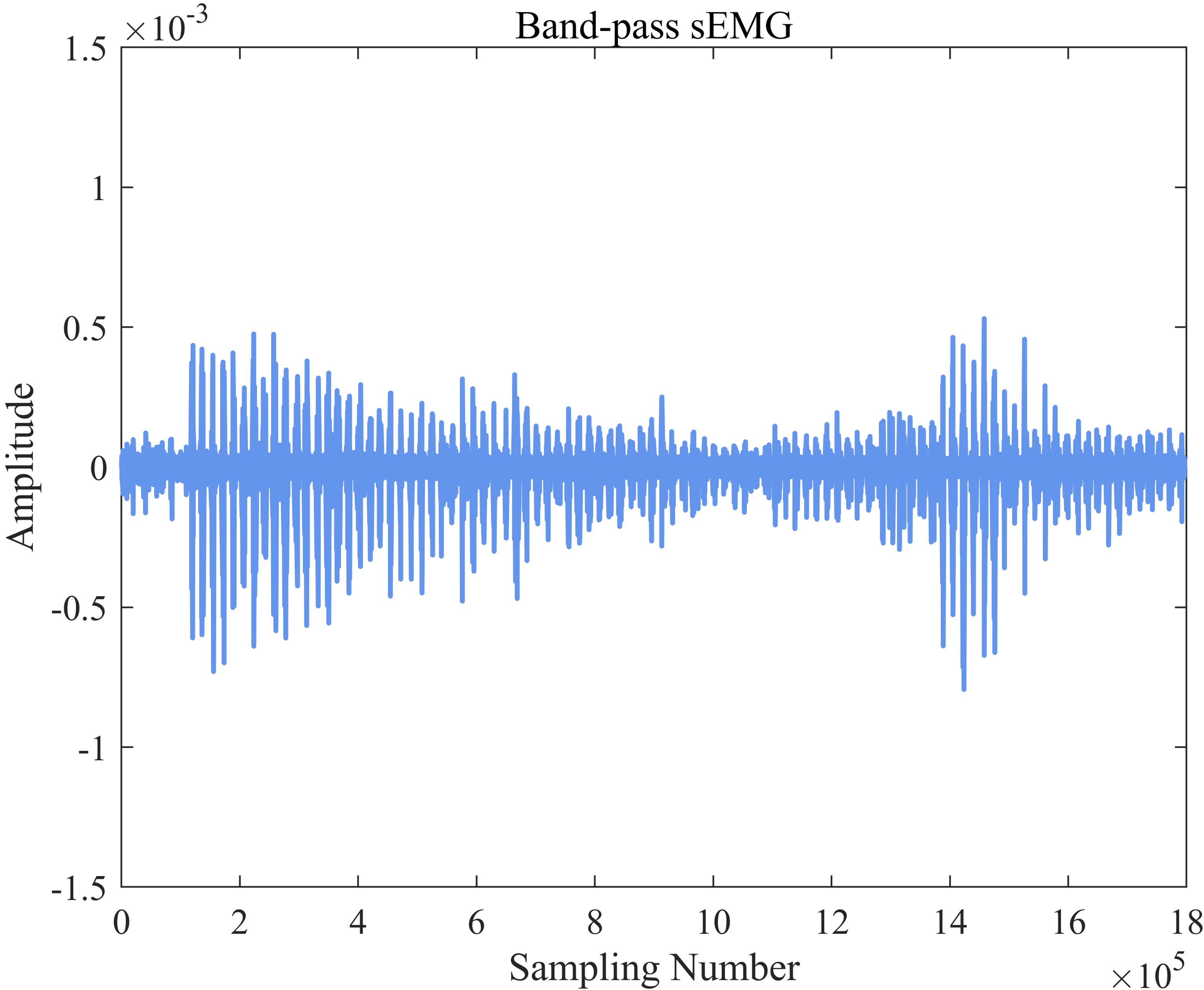}\includegraphics[width=0.16 \textwidth]{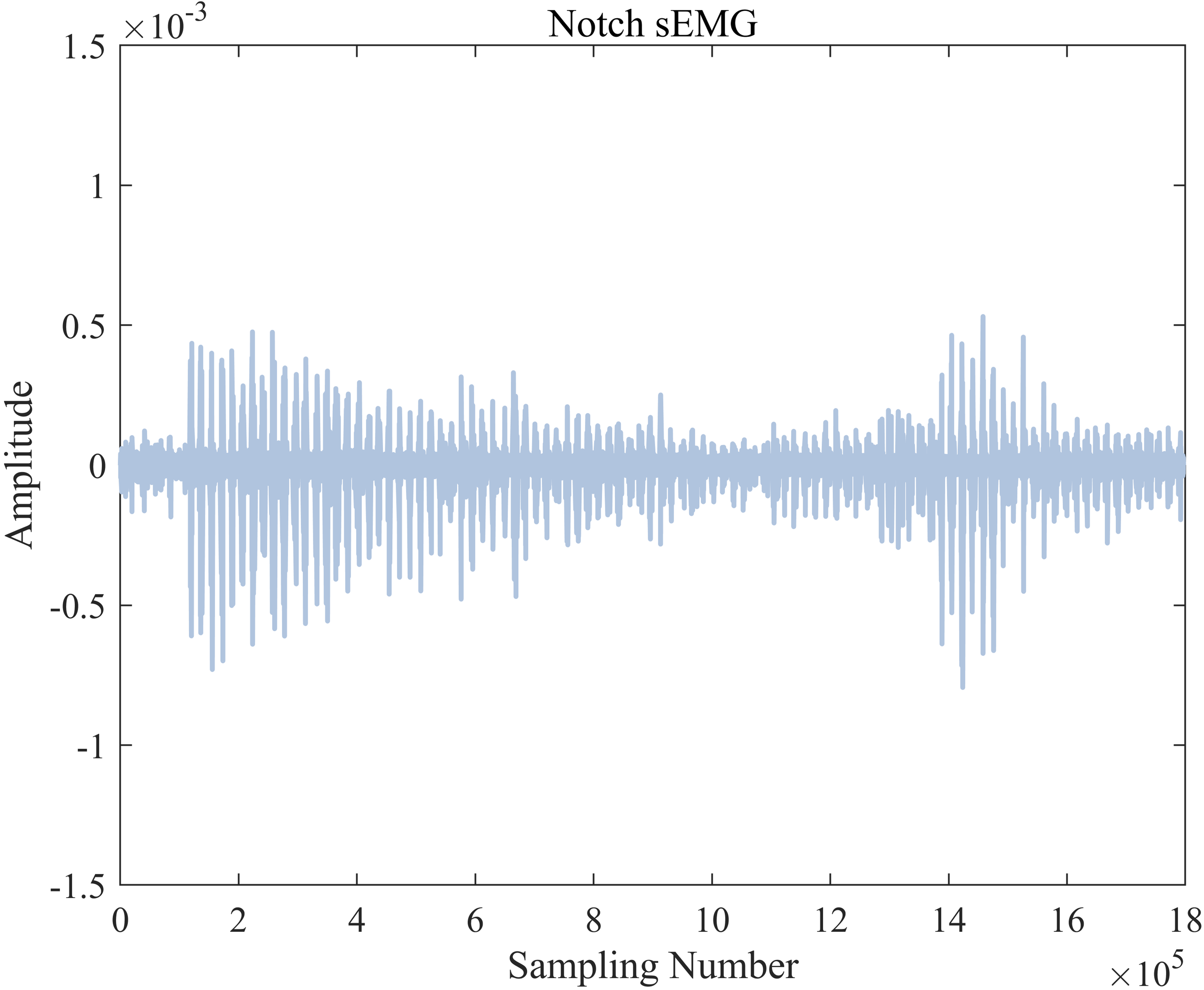}\\
	\includegraphics[width=0.16 \textwidth]{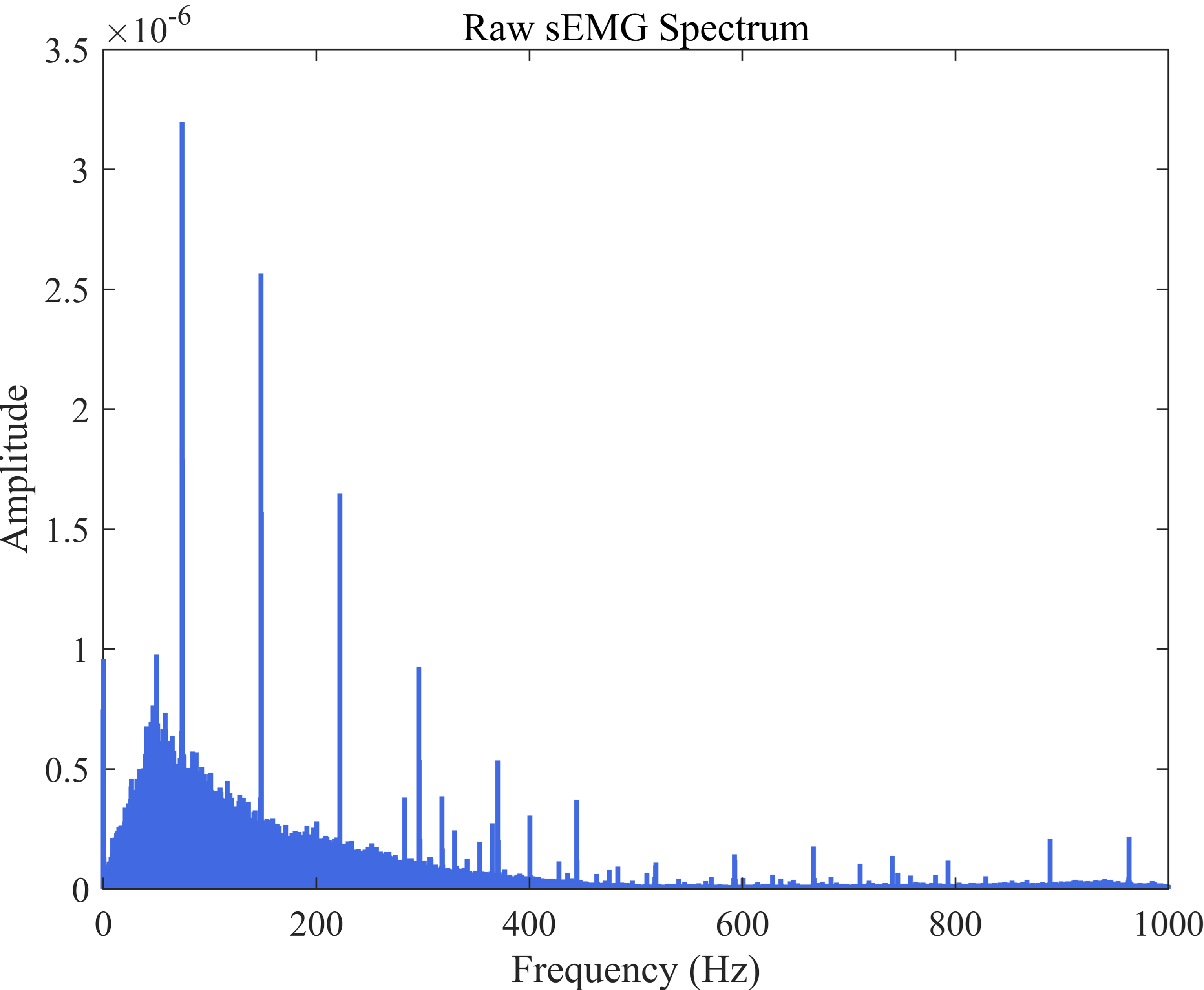}\includegraphics[width=0.16 \textwidth]{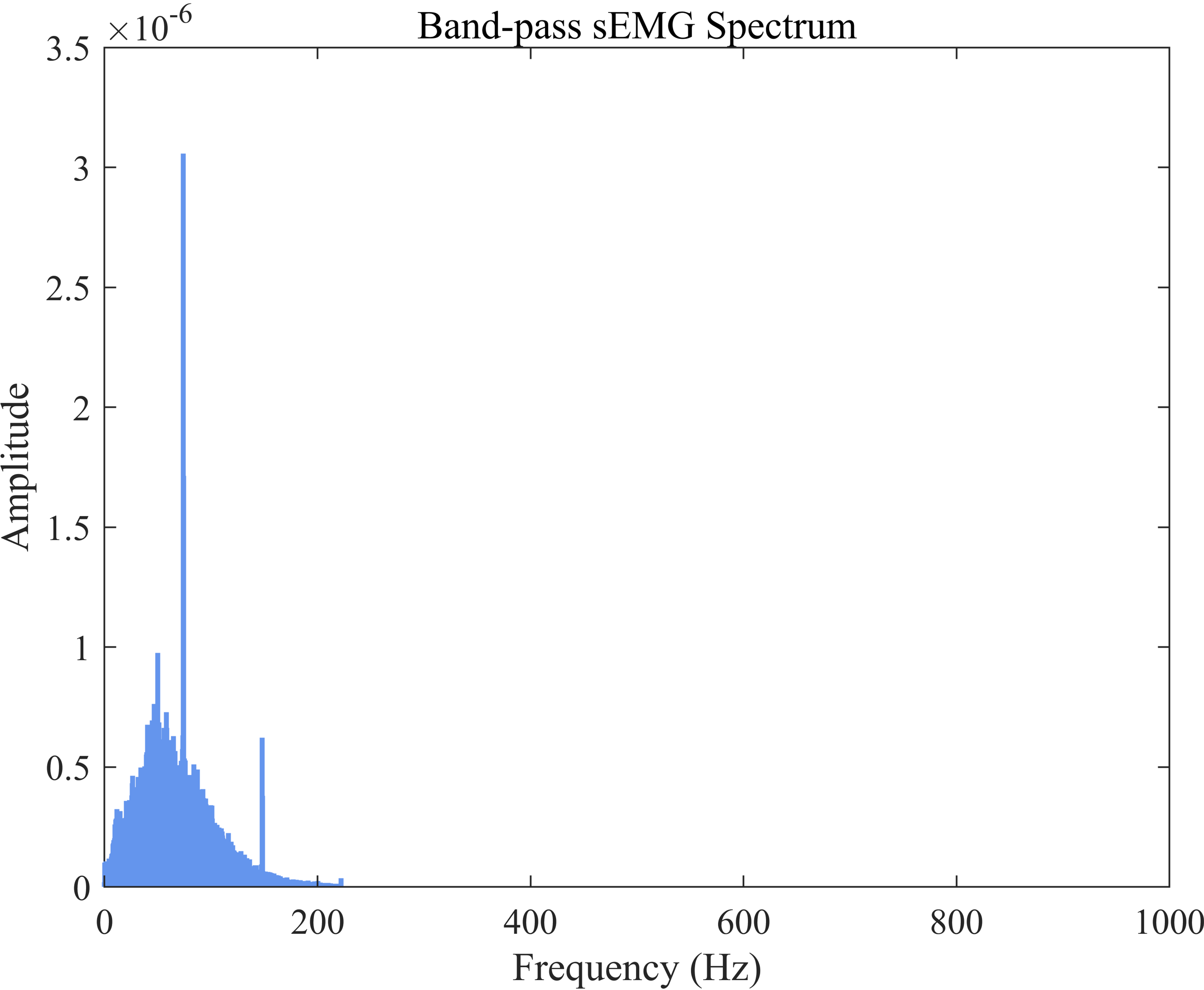}\includegraphics[width=0.16 \textwidth]{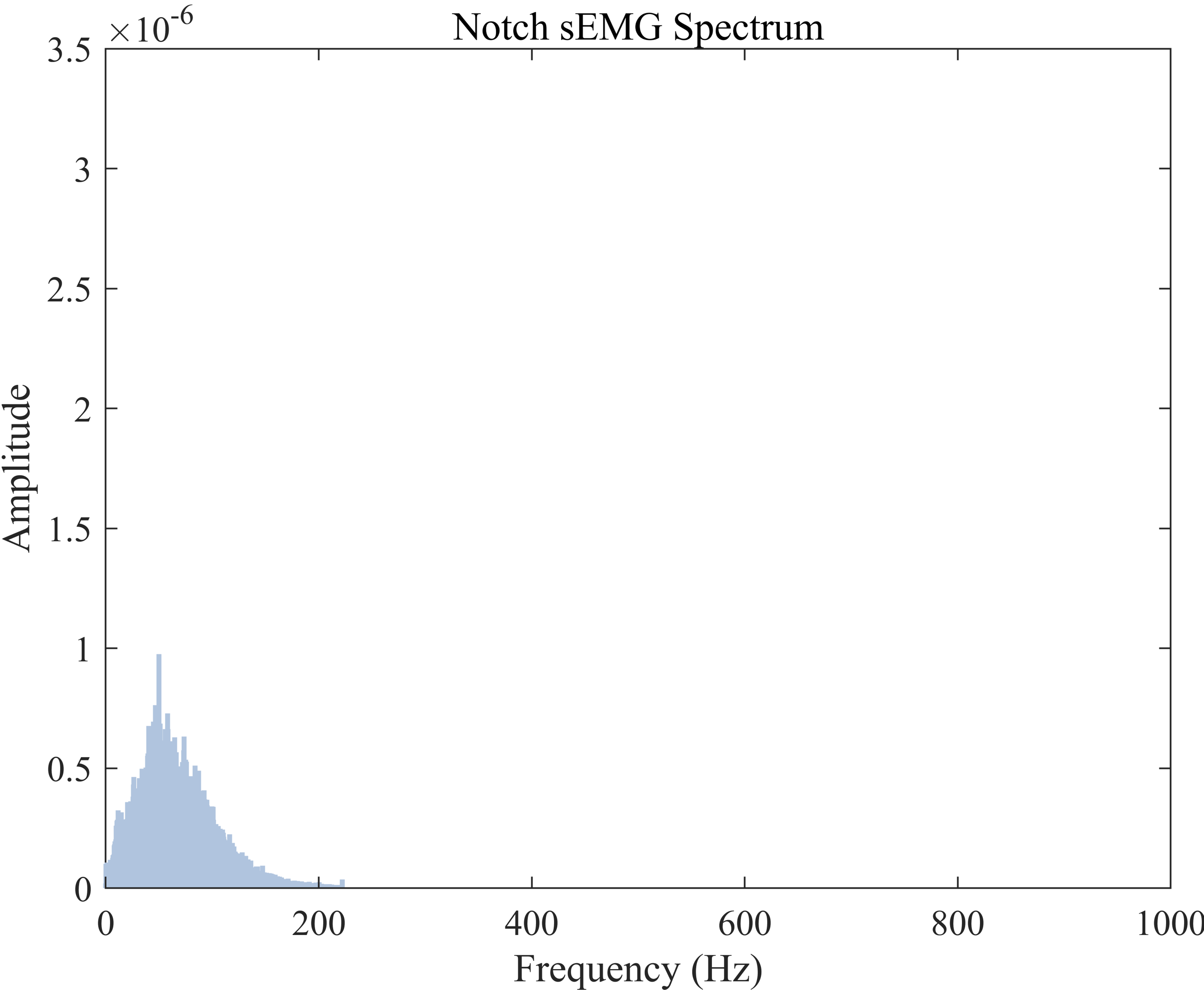}\\
	\raggedright\footnotesize{\ \ \quad\qquad (a) Original \ \ \qquad (b) Band-pass filtering \ \ \quad (c) Notch filtering}
	\caption{Comparison before and after filtering. Time domain signal (top) and spectrum (bottom)}
	\label{fig3}       
\end{figure}

According to existing studies, the most useful information in surface EMG is distributed in the frequency band of 0Hz-500Hz \cite{ding2018semg}, in which the main energy is concentrated in 20Hz-200Hz\cite{chen2020hand}. Therefore, we first adopted 20Hz-200Hz bandpass filtering, and used a 5-order Butterworth filter to perform band-pass filtering operation on 12-channel sEMG data. As shown in Fig. \ref{fig3} (b), the designed bandpass filter successfully filtered out low-frequency and high-frequency signals and improved the signal-to-noise ratio of the data, but there were still serious power frequency interference in the data. In the experimental environment of the current data set, the power frequency interference is 74hz, and more is 148hz, so notch filters are used to filter out the power frequency interference, as shown in Fig. \ref{fig3} (c). Finally, the filtered signal is standardized to improve the reliability of data analysis and facilitate the subsequent model training.
\begin{table}[!t]
	\caption{Dataset cross-validation grouping}
	\label{tab:1}
	\centering
	{ 
		\begin{tabular}{ccc} 
			\toprule
			{Group number} & {Training set} & {Testing set}  \\ 
			\hline
			1                     & Movements 2,4,5,6     & Movements 1,3         \\
			2                     & Movements 1,3,4,6     & Movements 2,5         \\
			3                     & Movements 1,2,3,5     & Movements 4,6         \\
			\bottomrule
	\end{tabular}}
\end{table}

\subsection{Feature Extraction}
\begin{figure}[!t]
	\centering
	\includegraphics[width=0.4 \textwidth]{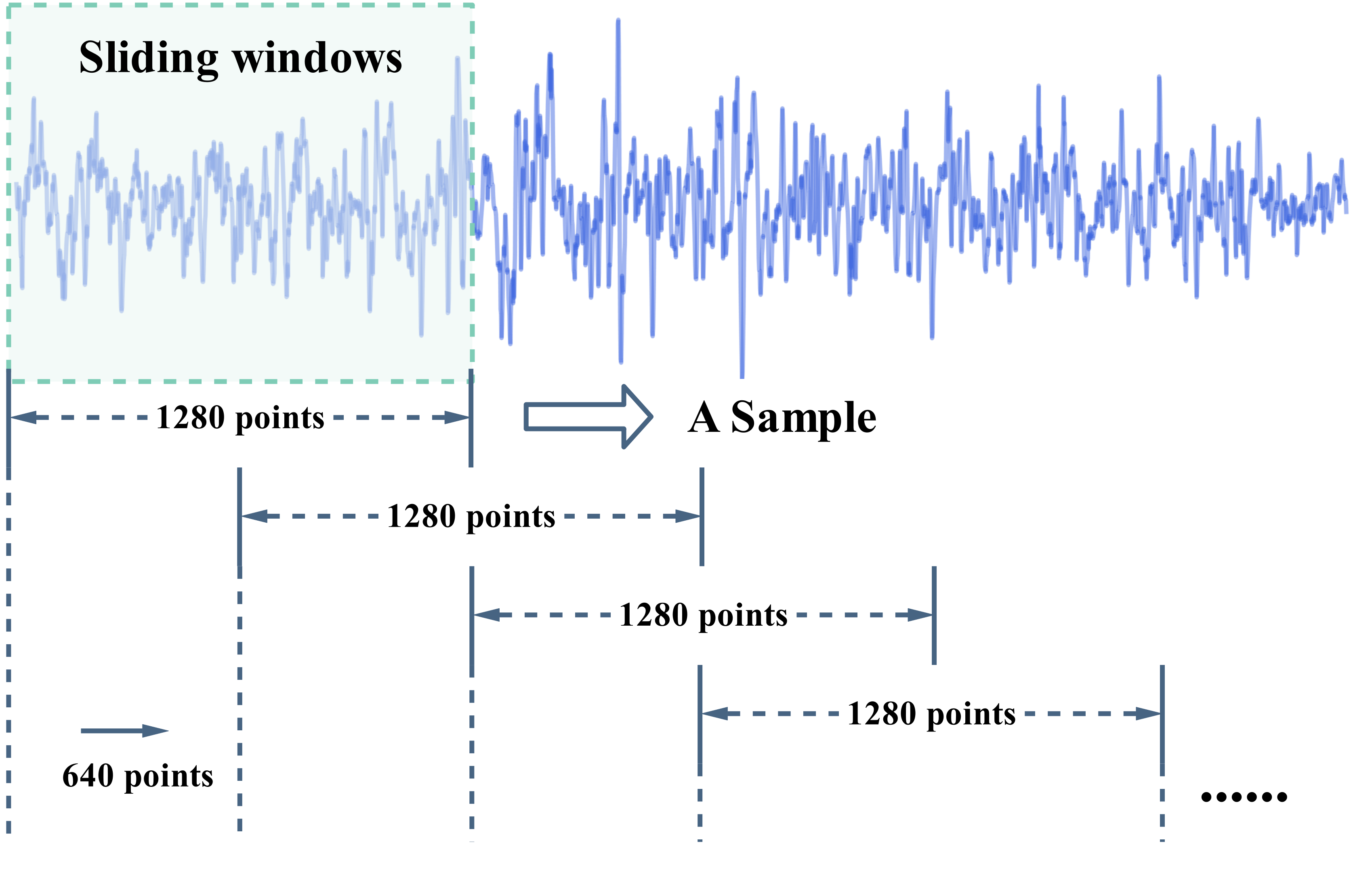}
	\caption{Sample diagram of sliding window segmentation}
	\label{fig4}       
\end{figure}

There is a certain relationship between the signal sampling frequency and the signal length of feature extraction. Generally speaking, the longer the signal length, the higher the frequency resolution, and the better the frequency component of the signal can be reflected. However, the signal length should not be too long, otherwise the time resolution will be reduced, and the time-varying characteristics of the signal cannot be captured \cite{cote2017transfer}. Therefore, in order to take into account the resolution of time domain features and frequency domain features, as shown in Fig. \ref{fig4}, data of 640 milliseconds is selected as a sample point. At a sampling rate of 2000hz, there are 1280 sample points, and the step length of the sliding window is 160 milliseconds, that is, 640 sample points.

Then feature extraction is carried out on these samples, because sEMG is a kind of time series signal, in which time domain features are strongly correlated with workers in the industry, so extracting time series features is an essential task in the feature engineering of gesture recognition \cite{tam2021intuitive}. In this paper, six features are selected as the feature set of time-domain features, including mean (MEAN), variance (VAR), average absolute voltage (MAV), zero crossing times (ZCR), root mean square (RMS) and waveform length (WL), and the expression is as follows
\begin{equation}\label{eq:1}
	\mu=\frac{1}{\mathrm{N}}\sum_{\mathrm{n=1}}^{\mathrm{N}}\mathrm{s(n)}\ ,
\end{equation}
\begin{equation}\label{eq:2}
	\sigma^{2}=\frac{1}{\mathrm{N-1}}\sum_{\mathrm{n=1}}^{\mathrm{N}}(s(\mathrm{n})-\mu)^{2} ,
\end{equation}
\begin{equation}\label{eq:3}
	\mathrm{MAV=\frac{1}{N}\sum_{n=1}^{N}|s(n)|},
\end{equation}
\begin{equation}\label{eq:4}
\begin{array}{c}
	\mathrm{ZCR}={\frac{1}{2N}}\sum_{\mathrm{n=1}}^{\mathrm{N}}\left|\mathrm{sgn}\big(\mathrm{s(n)}\big)-\mathrm{sgn}\big(\mathrm{s(n-1)}\big)\right| \\
	\\
\begin{cases}\:\mathrm{sgn\big(s(n)\big)=1,s(n)>0}\\\:\mathrm{sgn\big(s(n)\big)=0,s(n)=0}\\\:\mathrm{sgn\big(s(n)\big)=-1,s(n)<0}_{\leftrightarrow}\end{cases}
\end{array},
\end{equation}
\begin{equation}\label{eq:5}
	\mathrm{RMS}=\sqrt{\frac{1}{\mathrm{N}}\sum_{n=1}^{\mathrm{N}}\mathrm{s(n)}^2},
\end{equation}
\begin{equation}\label{eq:6}
	\mathrm{WL=\sum_{n=1}^{N}|s(n)-s(n-1)|}, 
\end{equation}
Where, $s(n)$ represents the size of surface EMG at the n-th point on the time series; $N$ is the number of surface EMG sampling points in the time series.

In addition, we also chose short-time Fourier Transform (STFT) to calculate the power spectral density of sEMG as the frequency domain feature. Specifically, the surface EMG data from 20hz to 200hz were divided into 10 frequency bands, and the power spectral density within the 10 frequency bands were calculated respectively, that is, 10 frequency domain features were extracted in a single channel as the features of fine gesture classification. Surface EMG data has 12 channels, so the total frequency domain cross section features are 12*10, and a total of 120 features. The expression is
\begin{equation}\label{eq:7}
	\mathrm{S(\omega,\tau)=F\big(m(t)w(t-\tau)\big)=\int_{-\infty}^{\infty}m(t)\omega(t-\tau)e^{-j\omega t}dt},
\end{equation}
\begin{equation}\label{eq:8}
	\mathrm{PSD(\omega)=\sum_{\omega}|S(\omega,\tau)|^2}, 
\end{equation}
Where $m(t)$ represents the input signal, $w(t)$ is the window function, $\omega$ is the frequency, and $\tau$ is the center position of the window function. Power spectral density is calculated $S(\omega,\tau)$ in the sum of squares of different frequency range.

sEMG feature extraction is basically studied in the time domain and frequency domain, while there are few studies on spatial features. When people are doing different gestures, muscle electrical signals are associated with different muscles. Therefore, a phase-locking Value (PLV) \cite{samadani2018gated,simao2019emg} feature was proposed in this paper, which was used as the spatial feature of surface EMG to describe the Phase synchronization between muscle activities. Before calculating the phase-lock value, it is necessary to use the analytic signal based on Hilbert transform to determine the instantaneous phase. Independent samples of a channel signal data $x(t)$, its analytical signal is:
\begin{equation}\label{eq:9}
	z\left(f,t\right)=x\left(f,t\right)+iy\left(f,t\right), 
\end{equation}
\begin{equation}\label{eq:10}
	\phi\left(f,t\right)=\angle\left(f,t\right)=arctan\frac{y\left(f,t\right)}{x\left(f,t\right)}, 
\end{equation}
\begin{equation}\label{eq:11}
	\text{PLV(f)}=|\langle\mathrm{e^{i(\phi_m(f,t)-\phi_n(f,t)}\rangle}|, 
\end{equation}
where $\langle\cdotp\rangle$ denotes the desired value, $\phi_m\left(f,t\right)-\phi_n\left(f,t\right)$ denotes the phase difference between the $m$ and $n$ electrode channels at $f$ frequency; PLV takes the value in the range of $[0, 1]$, where $1$ denotes full phase synchronization and $0$ denotes no phase synchronization; and $i$ is an imaginary unit. After traversing the signal data of all channels in each independent sample, the phase-lock values between the signal data of each two channels in each independent sample are calculated respectively, and these results are used as the connection strength between the signal data of different channels to construct the channel connection matrix. That is, go through the data of 12 channels of a sample, and calculate the PLV value of each two channels according to the above way, so that 12*12 values can be calculated to form a 12*12 matrix. For the machine learning model, the input features need to be cross-section features, so the phase-locked value features are expanded into 144 features.

\subsection{Improved LightGBM Model}
LightGBM \cite{xue2023underwater} algorithm is a decision tree algorithm based on gradient lifting. It is optimized on the traditional GBDT algorithm in many aspects, improves the training speed and accuracy, and reduces the memory consumption. It is widely used in various fields.

\subsubsection{Optuna hyperparameter optimization.} There are many default parameters in LightGBM algorithm, and using default parameters for different tasks will reduce the performance of the model. How to search the optimal parameters for gesture recognition task becomes the key task of this study. In this paper, Optuna hyperparameter algorithm is combined with LightGBM algorithm, and the accuracy rate is set as the optimization direction of hyperparameter. The optimization parameter search of Optuna under LGBM algorithm is realized. The core of Optuna algorithm is a gradient-based unilateral sampling Bayesian optimization method, which uses the information of historical trial to establish the posterior distribution of the objective function through Gaussian process (GP), and then selects the next most promising hyperparameter combination according to the sampling strategy (such as expectation raise (EI)) \cite{wang2020semg}. In each iteration, only a part of the sample with the larger absolute gradient value is retained, while a part of the sample with the smaller absolute gradient value is discarded randomly, which can reduce the calculation amount while retaining important information. To compensate for the impact of discarded samples, GOSS also reweights the retained samples so that their gradient sum, sample size sum are the same as the original data. At the same time, Optuna algorithm uses a mutually exclusive feature bundling method, which can losslessly reduce the feature dimension and improve the training efficiency without affecting the performance of the model. For features that are mutually exclusive, that is, features that do not take non-zero values at the same time, they can be bundled together to form a new feature, which can reduce the number of features and save memory space. To avoid overlapping of bundled feature values, EFB also adds an offset to each feature so that their value ranges do not intersect. Algorithm \ref{alg:1} shows the pseudo-code for implementing Optuna hyperparameter optimization in LightGBM model training.

\begin{algorithm}[!t]
    \footnotesize
    \caption{: \footnotesize{Hyperparameter Optimization Using Optuna}}
    \begin{algorithmic}[1] \label{alg:1}
        \REQUIRE training data $train\_x$, $train\_y$, test data $test\_x$, $test\_y$.
        \ENSURE optimized model parameters $param$.
        \STATE \textbf{Initialize:} $param=fixed\_parameters\_dict$.
        \FOR{$parameter\_name$, $interval$ in $search\_parameters.items()$}
            \STATE $param[parameter\_name] = trial.suggest(parameter\_name, interval)$;
            \STATE Train the model with $train\_x$, $train\_y$ and $param$;
            \STATE Calculate accuracy on $test\_y$ as $test\_acc$;
            \STATE \textbf{return} $test\_acc$;
        \ENDFOR
        \STATE Create Optuna study to maximize accuracy;
        \STATE $study.optimize(objective, n\_trials=search\_rounds)$;
    \end{algorithmic}
\end{algorithm}

\subsubsection{Re-integration of LGBM models.} 
\begin{figure}[!t]
	\centering
	\includegraphics[width=0.46 \textwidth]{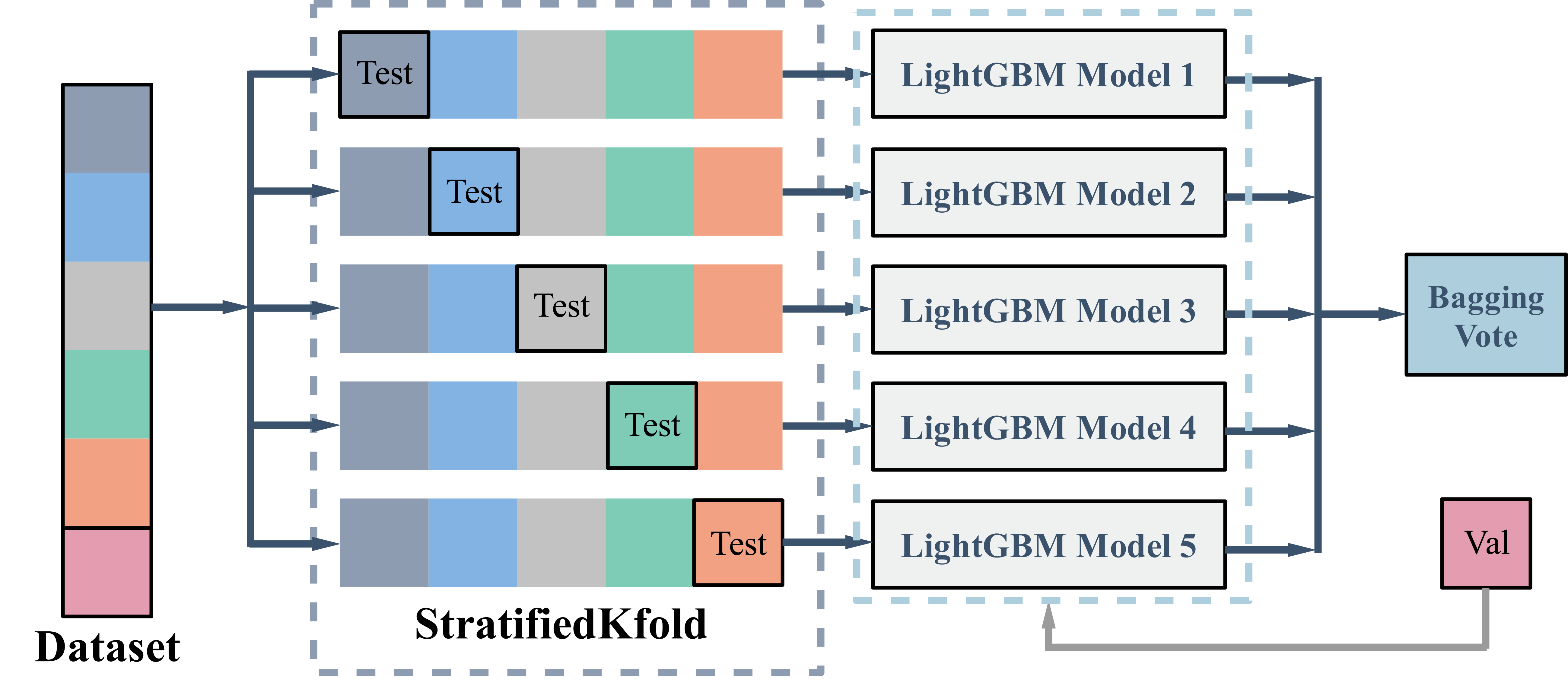}
	\caption{The schematic diagram of bagging LightGBM model}
	\label{fig5}       
\end{figure}

In the process of training, due to the uneven distribution of labels between the test set and the training set and the instability of the corresponding relationship, the accuracy will be too low. To solve this problem, the StratifiedKfold method \cite{tsinganos2019improved} is introduced in this study and combined with bagging to optimize the LGBM model. StratifiedKfold is a hierarchical cross-validation method used to evaluate the generalization ability of machine learning models. The basic idea is to divide the data set into k subsets according to the proportion of category labels, then select one subset each time as the test set, and the rest as the training set to conduct k times of model training and testing, and finally calculate the average value of k times of test results as the performance index of the model. The advantage of StratifiedKfold is that it ensures that each training and testing dataset reflects the class distribution of the original dataset, avoiding model bias due to data imbalance. The StratifiedKfold approach is suitable for classification problems, especially multi-classification problems, and can effectively evaluate the model's performance across different classes. Based on StratifiedKfold, this paper splits the training set into a 50-fold data set to train 5 LightGBM models, and then bagging 5 LightGBM models to integrate, which greatly improves the model's anti-overfitting ability. The schematic diagram of bagging LightGBM model is shown in Fig. \ref{fig5}.
\begin{algorithm}[!t]
    \footnotesize
    \caption{: \footnotesize{Improved Loss Function for sEMG Signal-Based Gesture Recognition}}
    \begin{algorithmic}[1] \label{alg:2}
        \REQUIRE predictions $preds$, data $data$, constant $K$, factor $fc$.
        \ENSURE gradient $grad$ and Hessian matrix $hess$.
        \STATE \textbf{Define:} $lambda\_c = calculate\_lambda\_c(K, fc)$;
        	\STATE $weighted\_ce\_loss = sum(lambda\_c * ce\_loss)$;
        	\STATE $grad = derivative(weighted\_ce\_loss)$;
        	\STATE $hess = second\_order\_derivative(weighted\_ce\_loss)$;
        	\STATE \textbf{return} $grad, hess$;
        \STATE Set loss function as $semgLoss$;
        \STATE $params['objective'] = semgLoss$;
        \STATE Train the model with $params$;
    \end{algorithmic}
\end{algorithm}

\begin{table*}[!t]
	\caption{Comparison of experimental results of improved LightGBM model}
	\label{tab:2}
	\centering
	{ 
		\begin{tabular}{cccccc} 
			\toprule
			movement & LightGBM & Optuna optimization & Bagging optimization & Loss optimization & Integration model  \\ 
			\hline
			1        & 70.52\%  & 83.01\%         & 78.67\%               & 87.59\%        & 88.20\%       \\
			2        & 84.81\%  & 93.71\%         & 90.26\%               & 84.83\%        & 96.63\%       \\
			3        & 85.63\%  & 93.77\%         & 89.63\%               & 85.89\%        & 96.94\%       \\
			4        & 92.01\%  & 95.10\%         & 91.80\%               & 92.16\%        & 98.01\%       \\
			5        & 76.27\%  & 83.81\%         & 79.27\%               & 76.24\%        & 87.08\%       \\
			6        & 80.70\%  & 82.86\%         & 79.34\%               & 80.76\%        & 89.16\%       \\
			7        & 88.07\%  & 97.27\%         & 93.98\%               & 88.01\%        & 97.05\%       \\
			8        & 85.50\%  & 96.51\%         & 92.93\%               & 85.62\%        & 98.40\%       \\
			9        & 70.81\%  & 79.99\%         & 76.40\%               & 71.21\%        & 82.82\%       \\
			10       & 75.15\%  & 83.75\%         & 79.85\%               & 75.12\%        & 86.66\%       \\
			11       & 75.61\%  & 82.43\%         & 78.30\%               & 75.32\%        & 86.12\%       \\
			12       & 65.88\%  & 74.78\%         & 70.88\%               & 70.19\%        & 78.63\%       \\
			13       & 60.29\%  & 69.54\%         & 65.56\%               & 72.02\%        & 78.05\%       \\
			14       & 84.22\%  & 93.76\%         & 90.36\%               & 84.24\%        & 96.37\%       \\
			15       & 80.86\%  & 86.91\%         & 83.03\%               & 81.52\%        & 91.30\%       \\
			16       & 90.49\%  & 98.35\%         & 94.33\%               & 90.33\%        & 98.62\%       \\
			17       & 71.17\%  & 76.61\%         & 73.52\%               & 71.06\%        & 85.30\%       \\
			18       & 79.93\%  & 84.70\%         & 80.55\%               & 80.52\%        & 89.64\%       \\
			mean     & 78.77\%  & 86.49\%         & 82.71\%               & 80.70\%        & \textbf{90.28\%}      \\
			\bottomrule
	\end{tabular}}
\end{table*}

\subsubsection{Improvement of loss function.}In the research process, it is found that some gestures belong to difficult samples, and the improvement of other ways will make the gestures with excellent performance better, while the accuracy of difficult gestures is not greatly improved. Based on such problems, this paper proposes an improved cross entropy loss function, as shown in the formula:
\begin{equation}\label{eq:12}
\begin{cases}L=-\dfrac{1}{N}\sum_{i=1}^{N}\sum_{c=1}^{M}\lambda_{c}y_{ic}\log(p_{ic})\\\lambda_{c}=ke^{1-f_{c}}(C)\end{cases}, 
\end{equation}
Where $N$ is the number of samples, $M$ is the number of categories, $C$ is the difficult subsample, $k$ is the gain coefficient, and $f_c$ is the frequency of the difficult subsample in the sample. Actions of different classes have different learning difficulties. Compared with cross-entropy loss function, for actions that are difficult to learn, increasing the weight of the loss function reduces the loss contribution of easily separable samples, thus making the model pay more attention to difficult samples and improving the classification performance of the model. The pseudo-code of the loss function is shown in Algorithm \ref{alg:2}.

\section{Results and Discussion}
\label{section3}
In this paper, 72 time domain features of 12-channel *6, 120 frequency domain features of 12-channel *10 and 144 spatial features are taken as inputs, and 18 kinds of gesture classification tasks are performed by using ordinary LightGBM algorithm and other commonly used algorithms, and the training results of each algorithm are compared.
\subsection{Normal Data Recognition}
\begin{figure}[!t]
	\centering
	\includegraphics[width=0.46 \textwidth]{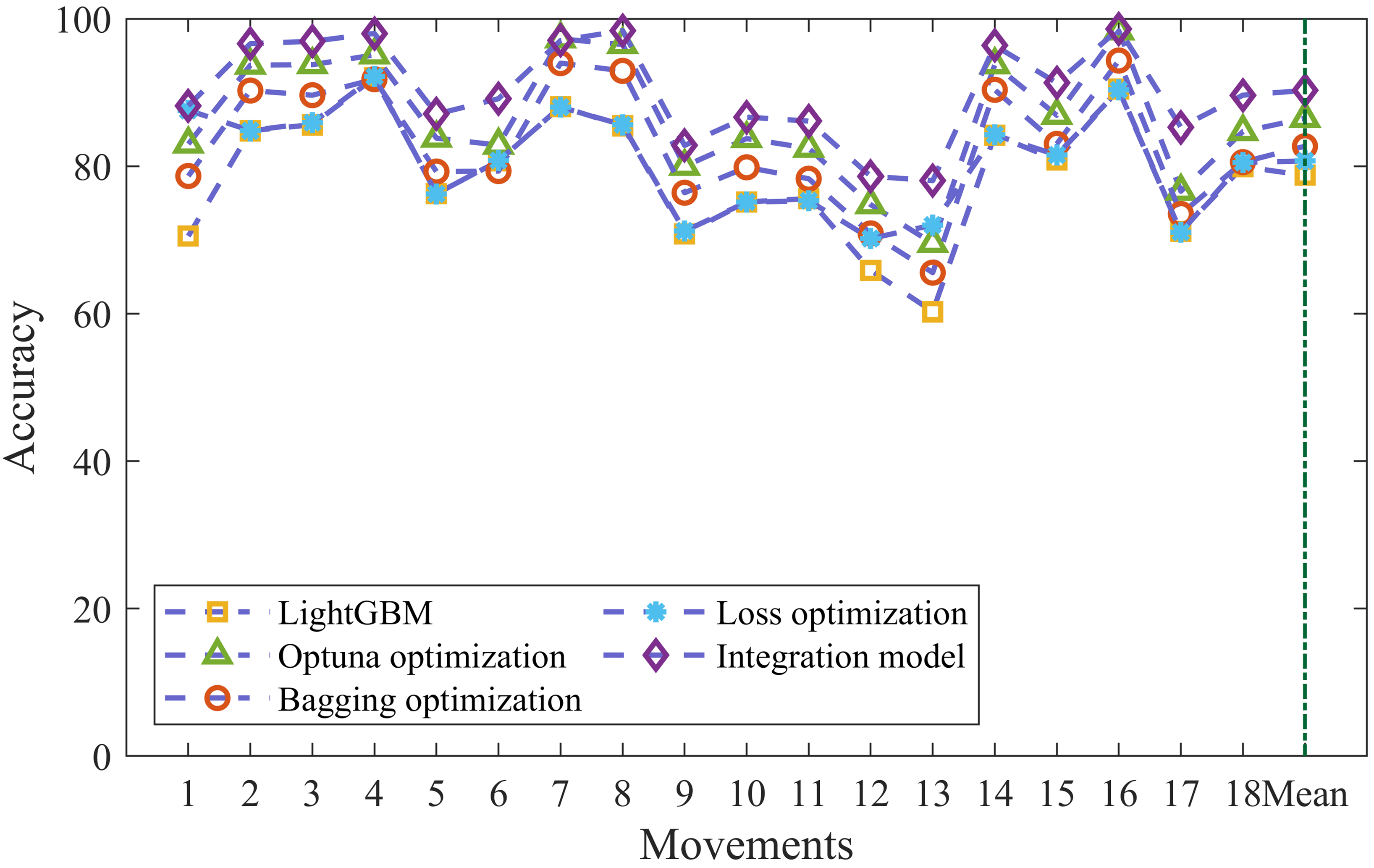}
	\caption{Optimize the comparison results of each stage}
	\label{fig6}       
\end{figure}

In this study, we used a variety of improvement and optimization schemes to improve the performance of the model, and compared whether the combined effect of multiple experimental methods produced positive feedback compared with the single optimization scheme. The experimental results are as shown in Table \ref{tab:2} and Fig. \ref {fig6}.

\begin{table*}[!h]
	\caption{Recognition rate and training time of different machine learning algorithms in gesture recognition}
	\label{tab:3}
	\centering
	{ 
	\begin{tabular}{ccccccc} 
		\toprule
		Movement      & LightGBM & Decision Tree & RF      & KNN     & XGBoost  & Ours\\ 
		\hline
		1             & 70.52\%  & 64.75\%       & 58.52\% & 29.93\% & 68.45\% & 88.20\%  \\
		2             & 84.81\%  & 67.54\%       & 81.47\% & 50.53\% & 84.97\% & 96.63\%  \\
		3             & 85.63\%  & 53.94\%       & 81.94\% & 49.68\% & 83.95\% & 96.94\%  \\
		4             & 92.01\%  & 60.17\%       & 87.47\% & 50.44\% & 88.53\% & 98.01\%  \\
		5             & 76.27\%  & 44.24\%       & 76.03\% & 40.81\% & 74.74\% & 87.08\%  \\
		6             & 80.70\%  & 45.42\%       & 80.26\% & 43.83\% & 77.15\% & 89.16\%  \\
		7             & 88.07\%  & 60.18\%       & 86.04\% & 67.02\% & 86.67\% & 97.05\%  \\
		8             & 85.50\%  & 59.42\%       & 86.08\% & 62.51\% & 82.80\% & 98.40\%  \\
		9             & 70.81\%  & 39.21\%       & 72.98\% & 40.79\% & 68.09\% & 82.82\%  \\
		10            & 75.15\%  & 57.81\%       & 70.51\% & 62.58\% & 76.25\% & 86.66\%  \\
		11            & 75.61\%  & 45.17\%       & 77.70\% & 56.74\% & 72.65\% & 86.12\%  \\
		12            & 65.88\%  & 39.16\%       & 74.78\% & 45.87\% & 67.04\% & 78.63\%  \\
		13            & 60.29\%  & 30.94\%       & 68.90\% & 43.16\% & 61.08\% & 78.05\%  \\
		14            & 84.22\%  & 55.60\%       & 82.07\% & 70.57\% & 83.37\% & 96.37\%  \\
		15            & 80.86\%  & 53.95\%       & 81.13\% & 63.63\% & 80.40\% & 91.30\%  \\
		16            & 90.49\%  & 58.83\%       & 87.23\% & 73.75\% & 85.93\% & 98.62\%  \\
		17            & 71.17\%  & 49.69\%       & 70.88\% & 57.71\% & 68.77\% & 85.30\%  \\
		18            & 79.93\%  & 50.42\%       & 84.30\% & 66.50\% & 77.91\% & 89.64\%  \\
		mean          & 78.77\%  & 52.01\%       & 78.24\% & 54.22\% & 77.15\% & 90.28\%  \\
		Training time & 28.9s    & 9.5s          & 1m23.5s & 8.2s    & 9m30.6s & 40s      \\
		\bottomrule
	\end{tabular}}
\end{table*}

After Optuna hyperparameters, the recognition accuracy of each action was improved, and the average recognition rate was increased from 78.77\% to 86.49\%. After StratifiedKfold and bagging integration, the recognition accuracy of each action is improved compared with the recognition results of ordinary LGBM models, and the average recognition rate increases from 78.77\% to 82.71\%. Compared with the fine gesture recognition model trained by the improved loss function, the four difficult actions numbered 1, 9, 12 and 13 have a great improvement compared with the ordinary LGBM model, and the recognition rate of action numbered 1 has increased from 70.52\% to 87.59\%. The recognition rate of action number 9 increased from 70.81\% to 71.21\%, the recognition rate of action number 12 increased from 65.88\% to 70.19\%, the recognition rate of action number 13 increased from 60.29\% to 72.02\%, and the average recognition rate increased from 78.77\% to 80.70\%. From the experimental results, this study all has played a positive role in improvement of Optuna super parameter optimization, bagging and integration solutions, loss function improvement on recognition rate increased, and the positive interaction between was, multiple improvement methods combined with the model in fine gesture recognition task achieves the optimum recognition rate of 90.28\%.

\begin{figure}[!t]
	\centering
	\includegraphics[width=0.5 \textwidth]{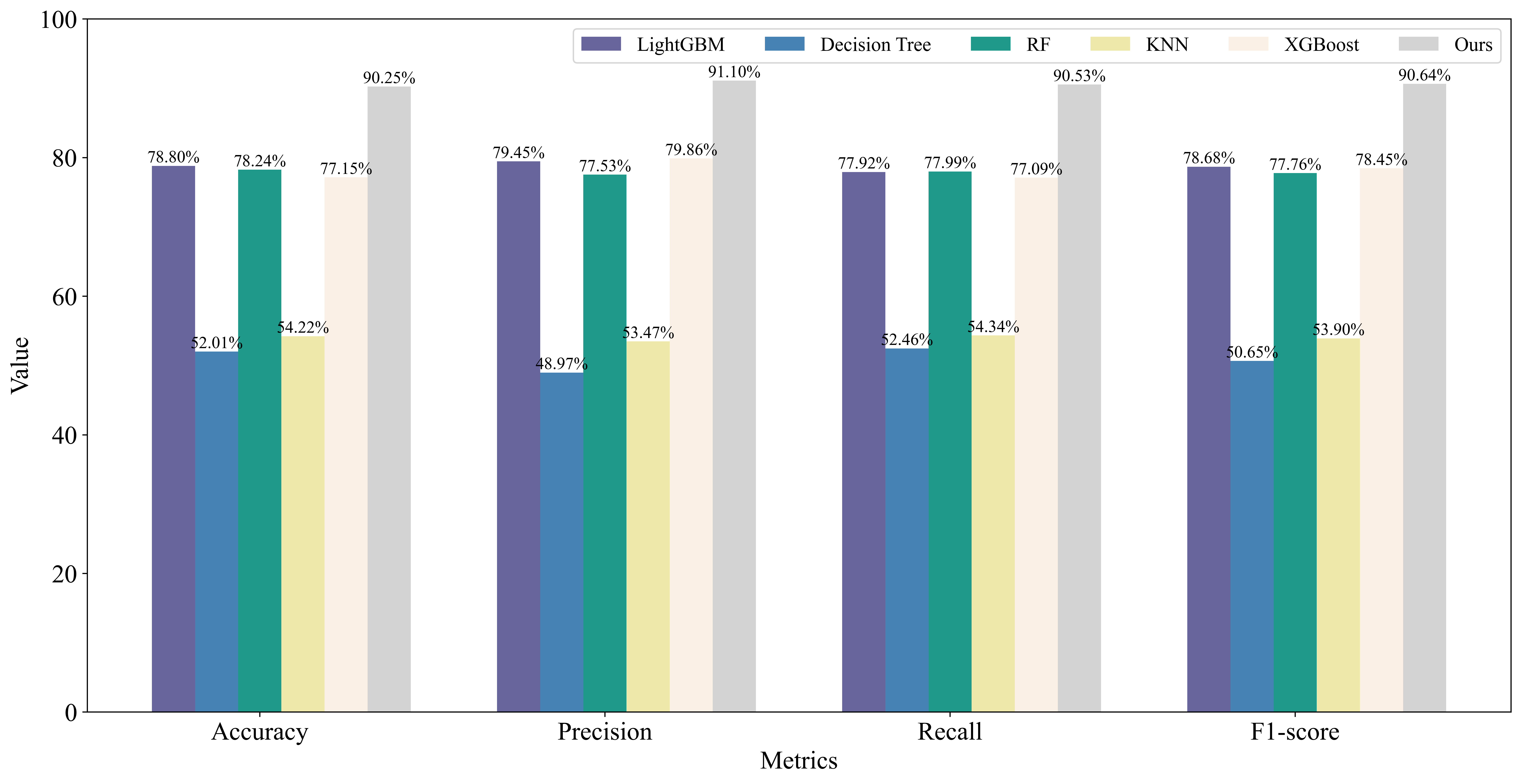}
	\caption{The comparison results of different algorithms}
	\label{fig7}       
\end{figure}

To prevent overfitting of the model, this paper sets accuracy as the condition of early-stop, and early-stop-rounds is set to 30 rounds. In terms of results, As shown in Table \ref{tab:3}, Table \ref{tab:4} and Fig. \ref{fig7}, LightGBM model, random forest model and XGBoost model in the integrated model all have good performance in recognition rate, while the decision tree and KNN model of single learner have poor performance. In terms of training time, XGBoost and random forest algorithm take longer time. The acceleration of LightGBM histogram algorithm makes LightGBM model achieve the best performance in training time, which lays the foundation for the realization of real-time online training gesture recognition system in the future of this research. Combining the accuracy and training time of 18 kinds of action classification, LightGBM algorithm achieves the best performance in fine gesture recognition.

\begin{table}
	\caption{The training results of different machine learning algorithms on fine gesture recognition}
	\label{tab:4}
	\centering
	\resizebox{\linewidth}{!}{ 
	\begin{tabular}{ccccccc} 
		\toprule
		Metrics   & LightGBM & Decision Tree & RF      & KNN     & XGBoost & Ours     \\ 
		\hline
		Accuracy  & 78.80\%  & 52.01\%       & 78.24\% & 54.22\% & 77.15\% & 90.25\%  \\
		Precision & 79.45\%  & 48.97\%       & 77.53\% & 53.47\% & 79.86\% & 91.37\%  \\
		Recall    & 77.92\%  & 52.46\%       & 77.99\% & 54.34\% & 77.09\% & 89.69\%  \\
		F1-score  & 78.68\%  & 50.65\%       & 77.76\% & 53.90\% & 78.45\% & 89.77\%  \\
		\bottomrule
	\end{tabular}}
\end{table}

\subsection{Disability Data Recognition}
\begin{figure}[!t]
	\centering
	\includegraphics[width=0.46 \textwidth]{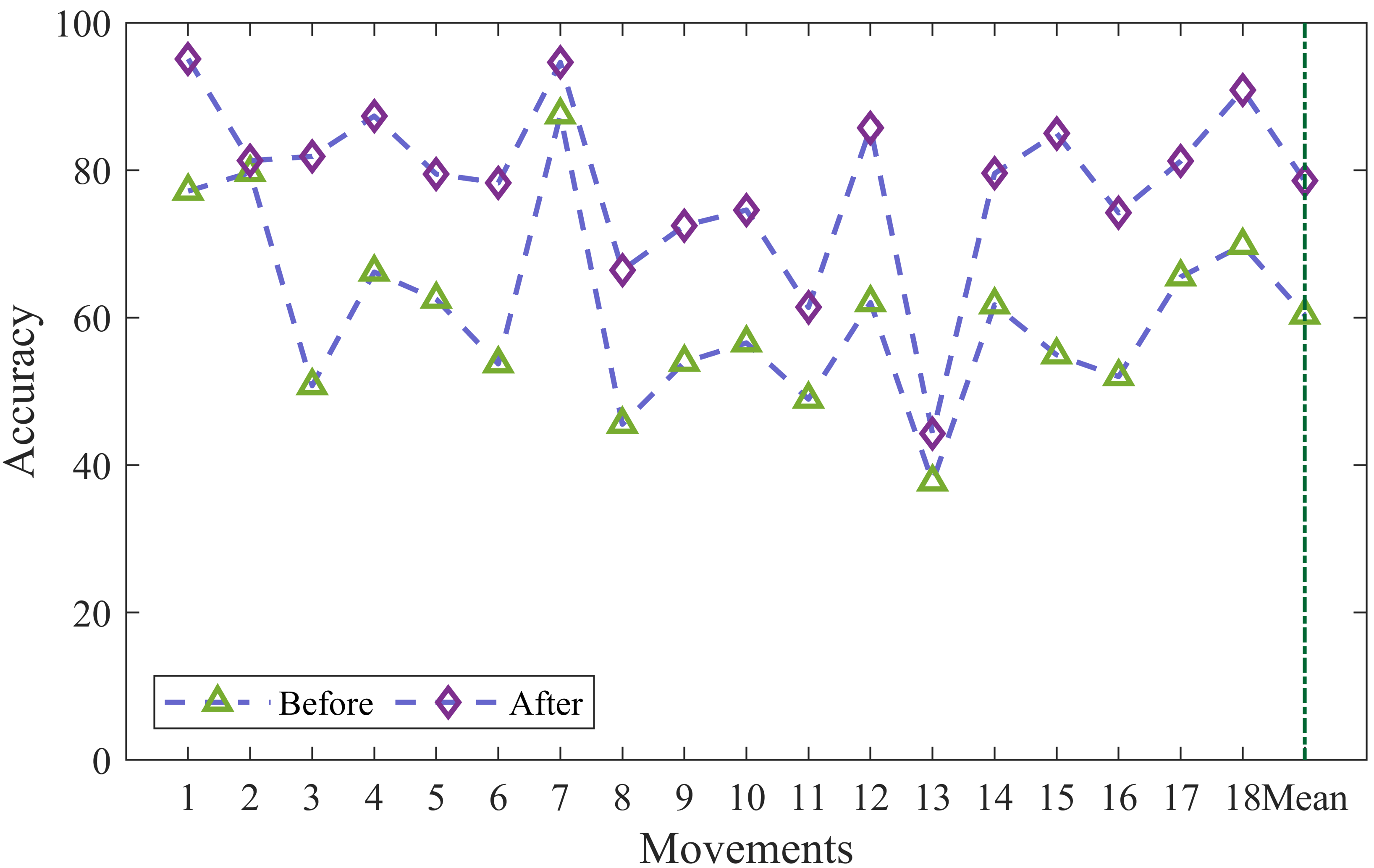}
	\caption{Comparison of results before and after transfer}
	\label{fig8}       
\end{figure}

The surface EMG data of the disabled is different from that of the normal people in terms of characteristics, but it is similar. The problem is that the data of the disabled is difficult to obtain and the amount of data is too small, which can be well solved by transfer learning \cite{betthauser2019stable}. There are 20 normal subjects and 2 disabled subjects in the data set of this study, and using the data set of 2 disabled subjects alone for fine gesture recognition will lead to the problem of too little training data. This study first uses the data of normal people to train the LightGBM tree model, and then saves the parameters of the model. The trained model parameters are used as initialization parameters for disabled data to start training. As a result, the precision of fine gesture recognition of surface EMG for disabled people has been greatly improved. The experimental results are shown in Table \ref{tab:5} and Fig. \ref{fig8}, 'Before' is the result of training the model directly with the data of the disabled, and 'After' is the result of training the data of the disabled with transfer learning. The transfer learning method improves the accuracy rate of fine gesture recognition of the data of the disabled by 18.19\%.

\section{Conclusion}
\label{section4}
This paper proposes an improved fine gesture recognition model based on LightGBM algorithm. Precision monitoring is used as the condition of early-stop to avoid the overfitting problem of LightGBM model, and Optuna is used to search for optimal hyperparameters to improve the performance of the model. The combination of StraitiedKold and bagging is used to reduce the influence of inconsistent distribution of training set and test set. The basic cross-entropy loss function is improved to make the model pay more attention to gestures that are difficult to recognize. The transfer learning method is used to realize the training and testing of small sample data of disabled people. Finally, the accuracy of the improved LightGBM model is compared with the results of the existing research. Through various improvement methods, the task of gesture recognition takes into account the training speed and recognition rate, and the accuracy rate of fine gesture recognition reaches 90.28\% in normal subjects and 78.54\% in disabled subjects. In future studies, we will pay more attention to the needs of disabled patients, and collect more data of disabled people to have a deeper understanding of their movement characteristics and needs, so as to tailor more appropriate assistance systems for them. 

\begin{table}[!t]
	\caption{Comparison of transfer learning experiment results}
	\label{tab:5}
	\centering
	\resizebox{\linewidth}{!}{ 
		\begin{tabular}{cccccc} 
			\toprule
			Movement & Before  & After & Movement & Before  & After  \\ 
			\hline
			1        & 77.16\% & 95.10\% & 10       & 56.57\% & 74.59\%  \\
			2        & 79.68\% & 81.29\% & 11       & 48.94\% & 61.39\%  \\
			3        & 50.77\% & 81.87\% & 12       & 62.03\% & 85.72\%  \\
			4        & 66.18\% & 87.34\% & 13       & 37.71\% & 44.27\%  \\
			5        & 62.50\% & 79.47\% & 14       & 61.73\% & 79.56\%  \\
			6        & 53.73\% & 78.28\% & 15       & 54.95\% & 85.00\%  \\
			7        & 87.50\% & 94.62\% & 16       & 52.00\% & 74.21\%  \\
			8        & 45.54\% & 66.45\% & 17       & 65.52\% & 81.23\%  \\
			9        & 53.95\% & 72.49\% & 18       & 69.81\% & 90.86\%  \\
			—        & —       & —       & mean     & 60.35\% & 78.54\%  \\
			\bottomrule
	\end{tabular}}
\end{table}

\appendices

%

\bibliographystyle{ieeetr}
\bibliography{myref}

%
%
%
%
%

\end{document}